
\input phyzzx.tex

{}~\hfill\vbox{\hbox{TIFR/TH/91-40}\hbox{September, 1991}}

\title{TWISTING CLASSICAL SOLUTIONS IN HETEROTIC STRING THEORY}

\author{S. F. Hassan and Ashoke Sen\foot{e-mail addresses:
FAWAD@TIFRVAX.BITNET, SEN@TIFRVAX.BITNET}}

\address{Tata Institute of Fundamental Research, Homi Bhabha Road,
Bombay 400005, India}

\abstract
We show that, given a classical solution of the heterotic string theory
which is independent of $d$ of the space time directions, and for which
the gauge field configuration lies in a subgroup that commutes with $p$ of
the $U(1)$ generators of the gauge group, there is an $O(d)\otimes O(d+p)$
transformation, which, acting on the solution, generates new classical
solutions of the theory.
With the help of these transformations we construct black 6-brane
solutions
in ten dimensional heterotic string theory carrying independent magnetic,
electric and
antisymmetric tensor gauge field
charge, by starting from a black 6-brane solution that carries magnetic
charge but no electric or antisymmetric tensor gauge field charge.
The electric and the magnetic charges point in different directions in the
gauge group.

\def\co{{\cal O}}
\def\to{\rightarrow}
\def\tpnew{\psi}
\def\cl{{\cal L}}
\def\ck{{\cal K}}
\def\cm{{\cal M}}
\def\cy{\check Y}
\def\ha{\hat A}
\def\ta{{\tilde a}}
\def\p{\partial}
\def\hy{\hat Y}
\def\ty{\tilde Y}
\def\odd{O(d)\otimes O(d)}

\def\r{\rangle}
\def\tg{\tilde G}
\def\hg{\hat G}
\def\tb{\tilde B}
\def\hb{\hat B}

\def\th{\tilde H}

\def\tr{\tilde R}
\def\tp{\chi}
\def\odo{O(d-1,1)\otimes O(d-1,1)}
\def\odp{O(d-1,1)\otimes O(d+p-1,1)}

\NPrefs
\def\define#1#2\par{\def#1{\Ref#1{#2}\edef#1{\noexpand\refmark{#1}}}}
\def\con#1#2\noc{\let\?=\Ref\let\<=\refmark\let\Ref=\REFS
         \let\refmark=\undefined#1\let\Ref=\REFSCON#2
         \let\Ref=\?\let\refmark=\<\refsend}

\define\DVV
R. Dijkgraaf, E. Verlinde and H. Verlinde, preprint PUPT-1252,
IASSNS-HEP-91/22.

\define\WITTEN
E. Witten,  Phys. Rev. {\bf D44} (1991) 314.

\define\ICHINOSE
I. Ichinose and H. Yamazaki, Mod. Phys. Lett. {\bf A4} (1989) 1509.

\define\MANDAL
G. Mandal, A.M. Sengupta and S.R. Wadia, preprint IASSNS-HEP-91/10.

\define\TSEYTLIN
A.A. Tseytlin, Mod. Phys. Lett. {\bf A6} (1991) 1721.

\define\ROCEK
M. Rocek, K. Schoutens and A. Sevrin, preprint IASSNS-HEP-91/14.

\define\BARDACKI
K. Bardacki, M. Crescimannu, and E. Rabinovici, Nucl.
Phys. {\bf B344} (1990) 344.

\define\CHS
C. G. Callan, J. Harvey and A. Strominger, Nucl. Phys. {\bf B359} (1991)
611.

\define\HOST
G. Horowitz and A. Strominger, Nucl. Phys. {\bf B360} (1991) 197.

\define\GIDST
S. Giddings and A. Strominger, preprint UCSBTH-91-35.

\define\ALOK
S. Khastgir and A. Kumar, Institute of Physics, Bhubaneswar preprint.

\define\GHS
D. Garfinkle, G. Horowitz and A. Strominger, preprint UCSB-TH-90-66.

\define\DGHR
A. Dabholkar, G. Gibbons, J. Harvey and F. Ruiz, Nucl. Phys. {\bf B340}
(1990) 33.

\define\DUFFLU
M.J. Duff and J. Lu, Nucl. Phys. {\bf B354} (1991) 141; Phys. Rev. Lett.
{\bf 66} (1991) 1402;preprint CTP/TAMU-29/91.

\define\CMP
C. G. Callan, R. C. Myers and M. Perry, Nucl. Phys. {\bf B311} (1988) 673.

\define\MYERS
R.C. Myers, Nucl. Phys. {\bf B289} (1987) 701.

\define\GIBBONS
G. Gibbons, Nucl. Phys. {\bf B207} (1982) 337;

\define\MAEDA
G. Gibbons and K. Maeda,
Nucl. Phys. {\bf B298} (1988) 741.

\define\VESA
H.J. de Vega and N. Sanchez, Nucl. Phys. {\bf B309} (1988) 552.

\define\MAZUR
P. Mazur, Gen. Rel. and Grav. {\bf 19} (1987) 1173.

\define\MYPE
R. C. Myers and M. Perry, Ann. Phys. {\bf 172} (1986) 304.

\define\LISTEIF
N. Ishibashi, M. Li and A.R. Steif, preprint UCSB-91-28.

\define\HORNE
J. Horne and G. Horowitz, preprint UCSBTH-91-39.

\define\NARAIN
K.S. Narain, Phys. Lett. {\bf B169} (1986) 41;
K.S. Narain, M.H. Sarmadi and E. Witten, Nucl. Phys. {\bf B279} (1987) 369.

\define\VAFA
C. Vafa, Private communications.

\define\VEW
A. Shapere and F. Wilczek, Nucl. Phys. {\bf B320} (1989) 669; A. Giveon,
E. Rabinovici and G. Veneziano, Nucl. Phys. {\bf B322} (1989) 167.

\define\MVE
K. Meissner and G. Veneziano, preprint CERN-TH-6138/91.

\define\VENEZIA
G. Veneziano, preprint CERN-TH-6077/91.

\define\SCALE
A.Sen, preprint IC/91/195 (TIFR-TH-91-35) (to appear in Phys. Lett. B).

\define\TWIST
A. Sen, preprint TIFR/TH/91-37.

\define\WILCZEK
A. Shapere, S. Trivedi and F. Wilczek, preprint IASSNS-HEP-91/33.

\define\EFR
S. Elitzur, A. Forge and E. Rabinovici, Preprint RI-143-90.

\define\BANE
I. Bars and D. Nemeschansky, Nucl. Phys. {\bf B348} (1991) 89.

\define\GMV
M. Gasperini, J. Maharana and G. Veneziano, preprint CERN-TH-6214/91.

\define\SENPOLY
A.~Sen,  Phys.~Lett. {\bf B241} (1990) 350.

\define\CECOTTI
S. Cecotti, S. Ferrara and L. Girardello, Nucl. Phys. {\bf B308} (1988)
436.

\define\DUFF
M. Duff, Nucl. Phys. {\bf B335} (1990) 610.

\define\SUPERGRAVITY
S. Ferrara, J. Scherk and B. Zumino, Nucl. Phys. {\bf B121} (1977) 393;
E. Cremmer, J. Scherk and S. Ferrara, Phys. Lett. {\bf B68} (1977) 234;
{\bf B74} (1978) 61;
E. Cremmer and J. Scherk, Nucl. Phys. {\bf B127} (1977) 259;
E. Cremmer and B. Julia, Nucl. Phys.{\bf B159} (1979) 141;
M. De Roo, Nucl. Phys. {\bf B255} (1985) 515; Phys. Lett. {\bf B156}
(1985) 331;
E. Bergshoef, I.G. Koh and E. Sezgin, Phys. Lett. {\bf B155} (1985) 331;
M. De Roo and P. Wagemans, Nucl. Phys. {\bf B262} (1985) 646;
L. Castellani, A. Ceresole, S. Ferrara, R. D'Auria, P. Fre and E. Maina,
Nucl. Phys. {\bf B268} (1986) 317; Phys. Lett. {\bf B161} (1985) 91.

\define\GAILLARD
M. Gaillard and B. Zumino, Nucl. Phys. {\bf B193} (1981) 221.

\define\NONPOL
M.~Saadi and B.~Zwiebach, Ann.~Phys. {\bf 192} (1989) 213;
T.~Kugo, H.~Kunitomo, and K.~Suehiro, Phys.~Lett. {\bf 226B} (1989) 48;

\define\GAUGEINV
T.~Kugo and K.~Suehiro, Nucl.~Phys. {\bf B337} (1990) 434.

\endpage

\chapter{INTRODUCTION}

It has been shown previously\con\VENEZIA\MVE\SCALE\TWIST\GMV\noc\ that in
any string theory, if we look
for solutions that are independent of $d$ of the space-time coordinates
$\hy^m$, then the space of such solutions has an $\odd$ or $\odo$ symmetry,
depending on whether the coordinates $\hy^m$ have Euclidean or Minkowski
signature.
(For definiteness, we shall assume from now on that the coordinates
$\hy^m$ have Minkowski signature).
Similar symmetries had been seen earlier in the context of supergravity
theories\SUPERGRAVITY, and invariance of classical equations of motion of
the two
dimensional $\sigma$ model under such transformations of the background
was shown in refs.\CECOTTI\DUFF\ following an earlier work of Gaillard and
Zumino\GAILLARD.
Under the action of this $\odo$ transformation a given solution
is in general mapped to an inequivalent solution.
In this paper we shall show that in heterotic string theory, if we look
for solutions that are independent of $d$ of the space-time coordinates,
and for which the background gauge field lies in a subgroup that commutes
with $p$ of the $U(1)$ generators of the gauge group, then the space of
such solutions has an $\odp$ symmetry.\foot{Invariance of the classical
equations of motion in the two dimensional $\sigma$-model under such
transformations of the background was discussed in ref.\CECOTTI.}
We shall also use this $\odp$ transformation to generate new classical
solutions of heterotic string theory starting from the known ones.

The plan of the paper is as follows.
In sect. 2 we present general arguments showing the existence of $\odp$
symmetry in the heterotic string theory for restricted class of
backgrounds of the type mentioned above.
In sect. 3 we study the manifestation of this symmetry in the low
energy effective field theory.
Section 4 contains application of this $\odp$ transformation on the known
solutions of the heterotic string theory, namely, the black p-brane
solution in ten dimensions carrying magnetic charge\HOST.
We show that by applying the $\odp$ transformation on this solution we can
generate a black $p$-brane solution carrying electric, magnetic and
antisymmetric tensor gauge field charge.
We summarise our results in sect. 5.

\chapter{$\odp$ SYMMETRY IN HETEROTIC STRING THEORY}

The origin of the $\odo$ symmetry discussed in refs.\VENEZIA-\GMV\ can be
traced to the fact\SCALE\ that if we restrict
to backgrounds that are independent of $d$ coordinates $\hy^m$,
then the interaction involving such backgrounds is governed by correlation
functions of vertex operators carrying zero $\hy^m$ momenta in the two
dimensional field theory of $d$ scalar fields $\hy^m$.
Such correlation functions factorise into products of
correlation functions in the holomorphic sector and the
antiholomorphic sector, each of which is separately invariant under the
Lorentz transformations involving the coordinates $\hy^m$.
In other words, these correlation functions have an $\odo$
symmetry.
As a result, the action involving $\hy^m$ independent background also has
an $\odo$ symmetry.
The manifestation of this symmetry in the context of low energy effective
field theory was found in refs.\VENEZIA\MVE\ (see also
refs.\SUPERGRAVITY\CECOTTI\DUFF) and was applied in
refs.\TWIST\GMV\ALOK\ to
generate new classical solutions in string theory from the known ones.

In heterotic string theory, besides the usual space-time coordinates we
also have $16$ internal coordinates which have only right moving
(anti-holomorphic) component
but no left moving (holomorphic) component.
If we consider backgrounds which are independent of $d$ of the space-time
coordinates, and carry zero momentum (and
winding number) in $p$ of the 16 internal coordinates, then the previous
argument
can be generalized easily to conclude that the space of such solutions
should have an $\odp$ symmetry.
For the sake of clarity, we shall now present this argument in some
detail.

The argument is best presented in the language of string field theory, so
let us first assume that there is an underlying string field theory that
governs
the dynamics of heterotic string theory, and that the vertices in this
string field theory are given in terms of correlation functions in the
conformal field theory describing the first quantised heterotic string
theory.
We consider heterotic string theory formulated in 10 dimensional
flat space time, although the argument can easily be generalised to the
case where some of the dimensions are replaced by an arbitrary (1,0)
super-conformal
field theory of the correct central charge.
Let $\{\Phi_a\}$ denote the set of basis states in the conformal field
theory (including the ghost part).
The string field is given by
$|\Psi\r =\sum_a \psi_a|\Phi_a\r$, with $\psi_a$'s as the dynamical
variables of the string field theory.
Then the general $N$ point interaction
vertex is given by $\psi_{a_1}\ldots \psi_{a_N} f_N(a_1,\ldots a_N)$, where
$f_N(a_1,\ldots a_N)$ denotes a quantity constructed out of an $N$ point
correlation function of some conformal transform of the fields
$\Phi_{a_1},\ldots \Phi_{a_N}$.\foot{Such a representation for the string
field theory action is known explicitly for the bosonic string
theory\con\NONPOL\GAUGEINV\SENPOLY\noc, but unfortunately not for the
heterotic or
the super-string theory.}
For describing the string field configurations which are independent of $d$
of the space-time coordinates (say $\hy^m$) and carry zero momentum in $p$
of the internal directions associated with the coordinates $\cy^R$ (say),
only those components $\psi_a$ will be non-zero, for which the
corresponding basis states $|\Phi_a\r$ have zero momentum in these $d$
space-time directions, and also zero momentum in these $p$ internal
directions.
A basis of such states in the conformal field theory
can be chosen in the form
$|\chi_{l}\r \otimes |\bar\chi_{\bar l}\r\otimes |\Phi'_{a'}\r$, where
$|\chi_l\r$ and $\bar\chi_{\bar l}\r$ are the basis states in the
holomorphic and the antiholomorphic sectors respectively of the conformal
field theory
described by the coordinates $\hat Y^m$ and $\check Y^R$, and
$|\Phi'_{a'}\r$ denote the basis of states in the conformal field theory
describing the
rest of the system.
The correlation functions involving these basis states on the sphere
factorise into
correlation functions involving the states $|\chi_l\r$, correlation
functions involving the states $|\bar\chi_{\bar l}\r$ and correlation
functions involving the states $|\Phi'_{a'}\r$.
Thus $f_N$ also has the
factorized form:
$$
f_N(l_1, \bar l_1, a'_1,\ldots l_N, \bar l_N, a'_N)
=f_N^{(1)}(l_1,\ldots l_N) f_N^{(2)}(\bar l_1,\ldots \bar l_N)
f^{(3)}_N(a'_1,\ldots a'_N)
\eqn\eexplicitone
$$
We now note that $f^{(1)}$ and $f^{(2)}$ are separately invariant under
$O(d-1, 1)$ and $O(d-1+p, 1)$ transformations respectively, which simply
correspond to
Lorentz transformations acting on the coordinates $\hy^m$ and $\cy^R$.
(Although the general correlation functions in the conformal field theory
describing the internal coordinates $\cy^R$ has no rotational symmetry due
to the fact that the torus is not invariant under an arbitrary rotation,
in the zero momemtum sector the correlation functions are completely
ignorant of the compactification of the coordinates $\cy^R$, and as a
result the correlation functions are not only symmetric under a rotation
among the coordinates $\cy^R$, but also the ones which mix $\cy^R$ and
$\hy^m$.)
This, in turn, implies that restricted to such backgrounds, the string
field theory action will have an $\odp$ symmetry, with the $O(d-1, 1)$
transformations acting on the index $l$ of $\psi_{l,\bar l, a'}$, and the
$O(d-1+p, 1)$ transformations acting on the index $\bar l$ of $\psi_{l,
\bar l, a'}$.

In the absence of a field theory describing the heterotic string theory,
the above arguments can be used to establish an $\odp$ symmetry of the
S-matrix elements when the external states are restricted to carry zero
momentum in $d$ of the space-time directions and zero charge under $p$ of
the U(1) generators of the gauge group.
Since the effective action of the theory is constructed from the S-matrix,
the $\odp$ symmetry must manifest itself as a symmetry of the effective
action also.
Note that this argument holds to all orders in the $\alpha'$ expansion,
since nowhere we had to assume that the momenta carried by the external
states in directions other than these $d$ directions are small.

Let us now see how this transformation acts on some specific components of
the string field.
Let $h_{\mu\nu}$, $b_{\mu\nu}$ and $a_{\mu R}$ denote the components of the
stri
   ng
field which couple to the graviton, the antisymmetric
tensor and the gauge field vertex operators respectively.
We shall choose normalizations such that $h_{mn}+b_{mn}$ couples to the
vertex operator $c\bar c\p \hy^m \bar\p\hy^n e^{i\tilde k .\tilde Y}$ and
$a_{mR}$ couples to the vertex
operator $c\bar c\p \hy^m\bar \p\cy^R e^{i\tilde k.\tilde Y}$, where
$\tilde
Y^\alpha$ denote the set of coordinates other than $\hy^m$.
Let $S$ and $R$ be the $O(d-1,1)$ and $O(d-1+p,1)$ matrices associated
with the Lorentz transformations involving the unbarred and the barred
indices respectively.
Then the $\odp$ transformation acts on these fields as:
$$
\pmatrix{ (h' + b') & a' \cr} = S \pmatrix{ (h + b) & a\cr} R^T
\eqn\etrsone
$$
where $\pmatrix{ (h + b) & a}$ is regarded as a $d \times (d+p)$ matrix.
Similarly, if $(h_{m\alpha}+b_{m\alpha})$ couples to the vertex operator
$c\bar c\p\hy^m\bar\p\ty^\alpha e^{i\tilde k.\ty}$, $(h_{\alpha
m}+b_{\alpha m})$
couples to the vertex operator $c\bar c\p\ty^\alpha\bar\p\hy^m e^{i \tilde
k.\ty}$, and $a_{\alpha R}$ couples to the vertex operator $c\bar
c\p\ty^\alpha
\bar\p\cy^R e^{i\tilde k.\ty}$, then these fields transform as,
$$\eqalign{
h'_{m\alpha}+b'_{m\alpha} =& S_{mn}(h_{n\alpha} + b_{n\alpha})\cr
h'_{\alpha m}+b'_{\alpha m} =& (h_{\alpha n}+b_{\alpha n}) R_{nm} +
a_{\alpha R} R_{Rm}\cr
a'_{\alpha R} =& (h_{\alpha n}+ b_{\alpha n}) R_{nR} + a_{\alpha S}
R_{SR}\cr
}
\eqn\etransformation
$$
Similar transformation laws can be derived for other fields as well, but
we shall not list them here.

Note that the $\odp$ `symmetry' described above holds for any background
that is
independent of $d$ of the space-time coordinates, and is neutral under $p$
of the U(1) subgroups of the gauge group.
This includes background massive fields as well.

\chapter{SYMMETRY OF THE LOW ENERGY EFFECTIVE ACTION}

Although the general argument guarantees the existence of an $\odp$
symmetry, realisation of this symmetry in terms of fields that appear in
the low energy effective action is somewhat non-trivial, since the
explicit relationship between the string field components $h_{\mu\nu}$,
$b_{\mu\nu}$ and the fields that appear in the low energy effective action
is not known.
What we would like to do now is to see how these
transformations can be realised in the context of low energy effective
field theory.
To do this we start with the low energy effective action of heterotic
string theory and rewrite it in such a way that its symmetry becomes
manifest.
The action is given by,
$$
S=-\int d^Dx\sqrt{\det G}e^{-\Phi}(\Lambda-R^{(D)}(G)+{1\over 12}
H_{\mu\nu\rho} H^{\mu\nu\rho}-G^{\mu\nu}\p_\mu\Phi\p_\nu\Phi
+{1\over 8} \sum_a F^a_{\mu\nu} F^{a\mu\nu}))
\eqn\eone
$$
where $G_{\mu\nu}$, $B_{\mu\nu}$, $A^a_\mu$ and $\Phi$ denote the graviton,
the antisymmetric tensor field, the gauge field, and the
dilaton, respectively, $F^a_{\mu\nu} = \p_\mu A^a_\nu -\p_\nu A^a_\mu
+ f^{abc} A_\mu^a A_\nu^b$,
$H_{\mu\nu\rho}=\p_\mu B_{\nu\rho}$ + cyclic permutations $ - (
\Omega^{(3)}_A)_{\mu\nu\rho}$, $R^{(D)}$
denotes the $D$ dimensional Ricci scalar, and $\Lambda$ is the
cosmological constant equal to $(D-10)/2$ for heterotic string.
$\Omega^{(3)}_A$ is the gauge Chern-Simons term given by
$(\Omega^{(3)}_A)_{\mu\nu\rho} = (1/4) (A^a_\mu F^a_{\nu\rho}$ +
cyclic permutations $ - f^{abc} A_\mu^a A_\nu^b A_\rho^c$).
The full effective action also involves Lorentz Chern Simons term,
but these are higher derivative terms and can be ignored to this order.
Let us now split the coordinates $X^\mu$ into two sets $\hat Y^m$ and
$\tilde Y^\alpha$ ($1\le m\le d$, $1\le \alpha\le D-d$) and consider
backgrounds independent of $\hat Y^m$.
Let us further concentrate on backgrounds where the gauge field background
lies in a subgroup that commutes with $p$ of the right moving $U(1)$
generators $\bar \p X^I$ associated with the internal coordinates $X^I$ of
the
heterotic string theory.
Let us denote the corresponding internal coordinates by
$\cy^R$ ($1\le R\le p$).
Let $\ta$ denote the gauge indices corresponding to the gauge generators
that commute with the $U(1)^p$ subgroup, and lie outside this subgroup.
Thus the allowed non-vanishing components of the gauge fields are
$A^\ta_\mu$ and $A^R_\mu$.

To begin with, we shall further restrict to background field
configurations for which
$G_{m\alpha}=B_{m\alpha}= A^R_\alpha = A^\ta_m =0$;
i.e. to backgrounds of the form $G=\pmatrix{\hg_{mn} & 0 \cr 0 &
\tg_{\alpha\beta}}$, $B=\pmatrix{\hb_{mn} & 0 \cr 0 & \tb_{\alpha\beta}}$,
$A^a_\mu =\{\tilde A^\ta_\alpha, \ha^R_m\}$.\foot{For such backgrounds,
the equations of motion obtained by varying the action with respect to the
field components that we have set to zero are satisfied identically.
Hence any invariance of the action for such restricted set of backgrounds
will also imply invariance of the complete set of equations of motion.}
Afterwards we shall see how to write down the transformation laws in the
general case when such restrictions are not there.
In this case, after an integration by parts, the action \eone\ can be
shown to take the form:
$$\eqalign{
S = - \int d^d\hy \int & d^{D-d} \ty \sqrt{\det \tg}
e^{-\tp}\Big[\Lambda-\tg^{\alpha\beta}\tilde
\p_\alpha\tp\tilde \p_\beta\tp -{1\over 32}\tg^{\alpha\beta} Tr(\tilde
\p_\alpha
M L \tilde \p_\beta M L)\cr
&-\tr^{(D-d)}(\tg)+{1\over
12}\th_{\alpha\beta\gamma}\th^{\alpha\beta\gamma}
+{1\over 8}\sum_{\ta} \tilde F^\ta_{\alpha\beta}\tilde
F^{\ta\alpha\beta}\Big]\cr
}
\eqn\etwo
$$
where,
$$
L=\pmatrix{\eta_d & 0\cr 0 & -\eta_{d+p}}
\eqn\etwoa
$$
$$
\tp=\Phi-\ln\sqrt{\det \hg}
\eqn\ethree
$$
and,
$$
M = \pmatrix{(K^T-\eta_d)\hg^{-1}(K-\eta_d) & (K^T -\eta_d) \hg^{-1}
(K+\eta_d) &
-(K^T-\eta_d) \hg^{-1} \ha\cr
(K^T+\eta_d)\hg^{-1}(K-\eta_d) & (K^T+\eta_d)\hg^{-1}(K+\eta_d) &
-(K^T+\eta_d)\hg^{-1}\ha\cr
-\ha^T\hg^{-1}(K-\eta_d) & -\ha^T \hg^{-1}(K+\eta_d) & \ha^T\hg^{-1}\ha\cr}
\eqn\efour
$$
Here $\eta_m$ denotes the $m$ dimensional Minkowski metric $diag
(-1,1,\ldots 1)$, $\ha$ is the matrix $\ha_{mR}\equiv \ha^R_m$, and,
$$
K = -\hb -\hg -(1/4) \ha\ha^T
\eqn\efive
$$
The action $\etwo$ is manifestly invariant under,
$$
M\to \Omega  M \Omega^T
\eqn\efivea
$$
$$
\tp\to\tp,~~~\tg_{\alpha\beta}\to\tg_{\alpha\beta},~~~
\tb_{\alpha\beta}\to \tb_{\alpha\beta},~~~~\tilde A^\ta_\alpha\to \tilde
A^\ta_\alpha
\eqn\esix
$$
where,
$$
\Omega = \pmatrix{S & \cr & R\cr}
\eqn\enewerone
$$
$S$ and $R$ being $O(d-1,1)$ and $O(d+p-1,1)$ matrices respectively.
At the linearised level, $\hg_{mn}=\eta_{mn}+h_{mn}$, $\hb_{mn}=b_{mn}$ and
$\ha^R_m = a_{mR}$, and the transformations given above agree with the
transformations of $h$, $b$ and $a$ given in eq.\etrsone.
The transformation law of $\Phi$ can also be shown to agree with the
linearised transformations\SCALE.
Also note that the action is in fact invariant under any $O(d, d+p)$
transformation generated by the matrices $\Omega$ satisfying $\Omega L
\Omega^T = L$, but the transformations outside the $\odp$ subgroup can be
shown to be pure gauge deformations\SCALE.

One way to derive the transformation laws given in eq.\efivea\ under the
$\odp$ transformation is as follows.\foot{This
argument was pointed out
by C. Vafa\VAFA.}
Let us imagine for
the time being that all the dimensions associated with the coordinates
$\hy^m$ have been compactified (the
effective action does not depend on whether these dimensions are compact
or not).
In that case the low energy effective field theory involving the moduli of
this compact space (together with the moduli associated with the internal
coordinates) is governed by the Zamolodchikov metric for these moduli,
which, in turn, is invariant under the $O(d, d+p)$ group introduced in
refs.\NARAIN.
Thus the action of the $\odp$ group on the various fields may be obtained
from the action of this $O(d, d+p)$ group on the moduli space.
This action, in turn, may be read out directly from the analysis of
ref.\VEW\ and is given by,
$$
M \to \Omega M \Omega^T
\eqn\enewone
$$
where $M$ is the same matrix as given in eq.\efour\ and $\Omega$ is an
$O(d, d+p)$ matrix which preserves the matrix diag($\eta_d, -\eta_{d+p}$).

Let us now consider the case where $G_{\alpha m}$, $B_{\alpha m}$ and
$A_{\alpha R}$ are non zero.
For simplicity, we shall assume here that the background gauge field is
Abelian, and belongs to the $U(1)^{16}$ subgroup of the gauge group.
Since this subgroup commutes with all the 16 U(1) generators, we can
take $p=16$.
In this case, we shall define a $(2D+16)\times (2D+16)$ matrix $\cm$ as:
$$
\cm = \pmatrix{(\ck^T-\eta_D)G^{-1}(\ck-\eta_D) & (\ck^T -\eta_D) G^{-1}
(\ck+\eta_D) &
-(\ck^T-\eta_D) G^{-1}  A\cr
(\ck^T+\eta_D)G^{-1}(\ck-\eta_D) & (\ck^T+\eta_D)G^{-1}(\ck+\eta_D) &
-(\ck^T+\eta_D)G^{-1} A\cr
- A^TG^{-1}(\ck-\eta_D) & - A^T G^{-1}(\ck+\eta_D) &  A^TG^{-1} A\cr}
\eqn\efournew
$$
where,
$$
\ck = -B -G -(1/4) AA^T
\eqn\efivenew
$$
The gauge index of $A$ now runs over all the 16
coordinates.
In this case, the full action can be expressed as,
$$\eqalign{
- \int d^d\hy \int & d^{D-d} \ty
e^{-\tpnew}\Big[\Lambda- G^{\alpha\beta}\tilde
\p_\alpha\tpnew\tilde \p_\beta\tpnew -{1\over 32} G^{\alpha\beta} Tr(\tilde
\p_\alpha
\cm \cl \tilde \p_\beta \cm \cl)\cr
&+ \tilde\p_\alpha\tpnew \tilde\p_\beta G^{\alpha\beta} - {1\over 2}
P_{(\alpha\beta)}^{\beta\alpha}
\Big]\cr
}
\eqn\etwonewer
$$
where,
$$
\cl=\pmatrix{\eta_D & 0\cr 0 & -\eta_{D+16}}
\eqn\etwoanewer
$$
and,
$$
\tpnew=\Phi-\ln\sqrt{\det  G}
\eqn\ethreenewer
$$
$P_{(\alpha\beta)}^{\beta\alpha}$ is
defined as follows.
We first define the matrices:
$$
V = \pmatrix{ \eta_D/\sqrt 2 & - \eta_D/\sqrt 2 & 0\cr
1/\sqrt 2 & 1/\sqrt 2 & 0 \cr
0 & 0 & 1\cr}
\eqn\eextraone
$$
$$
\check\cm = {1\over 2} V \cm V^T =\pmatrix{ G^{-1} & - G^{-1}\ck & G^{-1}
A/\sqrt 2 \cr
- \ck^T G^{-1} & \ck^T G^{-1} \ck & - \ck^T G^{-1} A/\sqrt 2\cr
A^T G^{-1}/\sqrt 2 & - A^T G^{-1} \ck/\sqrt 2 & A^T G^{-1} A /2\cr
}
\eqn\eextratwo
$$
$$
\check\cl = V \cl V^T = \pmatrix{ 0 & 1_D & 0 \cr 1_D & 0 & 0 \cr
0 & 0 & -1_{16}\cr}
\eqn\eextrathree
$$
We now define the matrices $P_{(\alpha\beta)}, \ldots, Z_{(\alpha\beta)}$
through the relations:
$$
(\p_{\alpha} \check\cm \check \cl (\check\cm -\check\cl)\check\cl \p_\beta
\check \cm) = \pmatrix{ P_{(\alpha\beta)} & Q_{(\alpha\beta)}
& R_{(\alpha\beta)}\cr S_{(\alpha\beta)} & T_{(\alpha\beta)} &
W_{(\alpha\beta)} \cr X_{(\alpha\beta)} & Y_{(\alpha\beta)} &
Z_{(\alpha\beta)}\cr}
\eqn\eextrafoura
$$
In the above, $P_{(\alpha\beta)}$ is a $D\times D$
matrix.
We now define $P_{(\alpha\beta)}^{\mu\nu}$ to
be the $\mu\nu$
component of this matrix.

The action given in eq.\etwonewer\ can be shown to be invariant under a
transformation of the form:
$$
\cm \to \Omega \cm \Omega^T,~~~~ \tpnew\to\tpnew
\eqn\eextrafour
$$
with,
$$
\Omega = \pmatrix{ 1_{D-d} &  &  & \cr
& S & & \cr & & 1_{D-d} & \cr & & & R \cr}
\eqn\eextrafive
$$
where $S$ and $R$ are the $O(d-1, 1)$ and $O(d+p -1,1)$ matrices discussed
previously (with $p=16$).
Note that when $G_{m\alpha}$, $B_{m\alpha}$ and $A_{\alpha R}$ are zero,
these transformation laws are identical to those given in eq.\enewone.
Also, these transformations reduce to the ones given in
eqs.\etrsone\ and \etransformation\ when $G_{\mu\nu}-\eta_{\mu\nu}$,
$B_{\mu\nu}$ and $A_{\mu R}$
are small and hence can be identified with $h_{\mu\nu}$, $b_{\mu\nu}$ and
$a_{\mu R}$ respectively.
The invariance of the action given in eq.\etwonewer\ under the symmetry
transformation given in eq.\eextrafive\ follows from the fact that
$G^{\alpha\beta}$ and $P_{(\alpha\beta)}^{\beta\alpha}$
remain invariant
under these transformations.

Before we conclude this section, let us remark that although the general
arguments of sect. 2 guarantee the existence of an $\odp$ `symmetry' of
the
string theory for appropriate backgrounds to all orders in $\alpha'$, it
does not guarantee that the transformation laws, when expressed in terms
of the fields $G_{\mu\nu}$, $B_{\mu\nu}$ and $\Phi$ will remain unchanged
when we include corrections that are higher order in $\alpha'$.
This is due to the fact that the functional relationship between the
string fields and the fields that appear in the effective field theory may
undergo modification when we include the effect of higher derivative
terms.
Evidence of such modification in the transformation laws has already been
seen\TSEYTLIN\DVV\SCALE.

\chapter{APPLICATION OF THE $\odp$ TRANSFORMATION}

We shall now apply the above transformations to known solutions of
heterotic string theory to generate new solutions.
In particular we shall take our starting solution to be the black six
brane solution of ref.\HOST\ carrying a magnetic charge.
(For related work see
refs.\con\CHS\GIDST\GHS\DGHR\DUFFLU\CMP\MYERS
\GIBBONS\MAEDA\VESA\MAZUR\MYPE\ICHINOSE\WILCZEK\LISTEIF\HORNE\noc.)
The solution is given by the following form of the metric and other
fields:
$$
ds^2 = -{(1-r_+/r)\over (1-r_-/r)} dt^2 + {dr^2\over (1-r_+/r) (1-r_-/r)}
+ r^2 d\Omega_2^2 +\sum_{i=1}^6 dX^i dX^i
\eqn\eseven
$$
$$
\Phi= -\ln (1-r_-/r) +\Phi_0
\eqn\eeight
$$
$$
F^1 = 2\sqrt 2 Q_M \epsilon_2
\eqn\enine
$$
where,
$d\Omega_2$ is the line element on a two sphere, and $\epsilon_2$ is the
volume form on the same two sphere.
$\Phi_0$, $r_+$ and $r_-$ are the three independent parameters labelling
the solution ($r_+>r_-$),
and $Q_M$ is the (quantised) magnetic charge carried by the black hole,
given by,
$$
Q_M = \sqrt{r_+r_-/2}
\eqn\eten
$$
For definiteness, we have taken the magnetic field to lie in the $U(1)$
subgroup generated by the first internal coordinate.
We shall now perform the $\odp$ transformation on this solution to
generate new solutions.
To this end, note that the solution is independent of the coordinate $t$
and also the six coordinates $X^i$, thus here $d=7$.
Furthermore, the presence of the magnetic field requires $A^1$ to have a
non-vanishing component tangent to the 2-sphere, thus if we want to satisfy
the condition $A_{\alpha R}=0$, we must exclude the direction 1 from the
set of directions $R$.
Although we have shown that this is not necessary, we shall first consider
this case.
Thus here $p=15$.
Although we can involve all the $7$ space-time coordinates, and all the
$15$ internal coordinates in the transformation, a general transformation
of this kind will generate solutions which will be related by rotation in
the external and/or internal space.\foot{For $E_8\times E_8$ heterotic
string theory rotation among  the 15 internal coordinates can generate
inequivalent field configurations since $O(16)$ is not a subgroup of the
gauge group.
But this only changes the direction of the gauge field in the
final solution without modifying the essential properties of the
solution.}
Thus the set of inequivalent field configurations are generated by taking
the appropriate `Lorentz transformations'among the coordinate $t$, one of
the space coordinates (say $X^1$) and one of
the internal coordinates ( say $\cy^2$).
The symmetry group in this case is $O(1,1)\otimes O(2,1)$.
The diagonal $O(1,1)$ subgroup corresponds to Lorentz transformation of
the solution in the $t-X^1$ space, we may fix a Lorentz frame by choosing
the matrix $S$ to be the identity matrix.
Thus we are left with the $O(2,1)$ matrix $R$ parametrized by the three
Euler angles.
A further reduction of the parameters may be made by noting that the
original solution is left invariant if we choose $R$ to be a rotation in
the $X^1-\cy^2$ plane.
Thus the general solution is obtained by taking $R$ to be a boost in the
$t - \cy^2$ direction followed by a boost in the $t - X^1$ direction:
$$\eqalign{ R= &
\pmatrix{\cosh\alpha_2 & \sinh\alpha_2 & 0\cr \sinh\alpha_2 &
\cosh\alpha_2 & 0\cr 0 & 0 & 1\cr}
\pmatrix{\cosh\alpha_1 & 0 & \sinh\alpha_1\cr 0 & 1 & 0 \cr \sinh\alpha_1
& 0 & \cosh\alpha_1\cr}\cr
}
\eqn\eeleven
$$
We can now calculate the transformed solution in a straightforward way
using eqs.\efivea, \esix.
The transformed solution is given by,
$$\eqalign{
ds^2 &= -{1\over 4(r- r_0)^2} (4 (r-r_+)(r-r_-)
-(r_+-r_-)^2\beta^2) dt^2 +
\beta {r_+ - r_-\over (r-r_0)} dX^1 dt \cr
&~~ +\sum_{i=1}^6 dX^i dX^i + {dr^2\over (1-r_+/r) (1-r_-/r)}
+ r^2 d\Omega_2^2 \cr
B_{t1} &=\beta {r_+ - r_-\over 2(r - r_0)}\cr
A^2_t &= \gamma {r_+-r_-\over (r-r_0)}\cr
A^2_1 &= 0\cr
F^1 &= 2\sqrt 2 Q_M\epsilon_2\cr
\Phi &= -\ln (1-r_0/r) +\Phi_0 \cr
}
\eqn\etwelve
$$
where,
$$\eqalign{
\gamma &= \sinh\alpha_1\cr
\beta &= \cosh\alpha_1 \sinh\alpha_2 \cr
r_0 &= {1\over 2} ((r_++r_-) - (r_+-r_-)\sqrt{1 +\beta^2 +\gamma^2})\cr
}
\eqn\ethirteen
$$
This solution is characterized by an electric field as well as an
antisymmetric tensor field strength, given by,
$$\eqalign{
F^2_{rt} &= \p_r A^2_t = -\gamma {(r_+ - r_-)\over (r-r_0)^2}\cr
H_{rt1} &= \p_r B_{t1} = -\beta {(r_+ - r_-)\over 2(r-r_0)^2}\cr
}
\eqn\efourteen
$$
Hence, besides carrying the magnetic charge, the new solution carries
both, electric and antisymmetric tensor gauge field charge, proportional
to $\gamma$ and
$\beta$ respectively.

Let us now discuss singularities of the solution \etwelve.
It can be easily seen that the matrix
$$
\pmatrix{G_{tt} & G_{t1}\cr G_{t1} & G_{11}}
$$
has zero eigenvalues at $r=r_+$ and $r= r_-$.
The component $G_{rr}$ has poles at precisely these values of $r$, as can
be seen from eq.\eseven.
These singularities represent coordinate singularities, and can be removed
by appropriate coordinate choice.
To see this, let us first define new coordinates $t'$, $X'$ and $\rho$
through the relations:
$$\eqalign{
t=&t'\cosh\theta - X'\sinh\theta\cr
X^1=&X'\cosh\theta -t'\sinh\theta\cr
r=&\rho+r_+\cr
}
\eqn\ecoordone
$$
where,
$$
\tanh\theta = {\beta\over 2} {r_+-r_-\over r_+-r_0}
\eqn\ecoordtwo
$$
In this coordinate system, the metric near $r=r_+$ takes the form:
$$\eqalign{
ds^2 =&-{2\cosh\theta\sinh\theta\over\beta (r_+-r_0)}\rho
dt^2 (1+\co(\rho)) \cr
& +{2 (r_+-r_0)\over\beta(r_+-r_-)}\Big[ 1-{\beta^2 (r_+-r_-)^2\over
4(r_+-r_0)^2} \Big]^2\cosh\theta\sinh\theta (dX')^2(1+\co(\rho))\cr
&-{2\rho\over r_+-r_0}\Big[ 1-{1\over 4}{\beta^2(r_+-r_-)^2\over
(r_+-r_0)^2} -{r_+-r_-\over r_+-r_0}\Big]\cosh\theta\sinh\theta
 dX' dt' (1+\co(\rho))\cr
&+ {(r_+)^2\over (r_+-r_-)\rho}(d\rho)^2(1+\co(\rho))
+(r_++\rho)^2(d\Omega_2)^2 +\sum_{i=2}^6 dX^idX^i\cr
}
\eqn\ecoordthree
$$
{}From this we see that the metric has a singularity at $\rho=0$.
This singularity is removed by defining new coordinates $u$, $v$ as,
$$
u=\sqrt\rho e^{at'},~~~v=\sqrt\rho e^{-at'}
\eqn\ecoordfour
$$
where,
$$
a=\sqrt{\cosh\theta\sinh\theta (r_+-r_-)\over 2 (r_+)^2 (r_+-r_0)\beta}
\eqn\ecoordfive
$$
In this coordinate system the metric takes the form:
$$\eqalign{
ds^2 =& {4(r_+)^2\over (r_+-r_-)} dudv +{2(r_+-r_0)\over\beta (r_+-r_-)}
\Big[ 1- {\beta^2(r_+-r_-)^2\over
4(r_+-r_0)^2}\Big]^2\cosh\theta\sinh\theta (dX')^2\cr
& +(r_+)^2 (d\Omega_2)^2 +\sum_{i=2}^6 dX^idX^i +\co(u, v)\cr
}
\eqn\ecoordsix
$$
{}From this we see that the metric is non-singular in this coordinate system
at $u=0$ or $v=0$.
It can also be seen easily that both the electric and the antisymmetric
tensor fields are non-singular at $r = r_+$ in the new coordinate
system.
Similar change of coordinates can also be carried out near $r=r_-$ to show
that this
also represents a coordinate singularity.\foot{In this case we again look
for a coordinate transformation of the form $t=t''\cosh\phi
-X''\sinh\phi$, $X^1=X''\cosh\phi -t''\sinh\phi$, $r=r_-+\rho'$ as in
eq.\ecoordone, so as to bring the metric in the standard form near the
singular surface.
It can be easily seen that if $|\beta|<\gamma^2/2$, then it is possible to
find a $\phi$ ($\tanh\phi=\beta (r_+-r_-)/2(r_--r_0)$) for which
$G_{t''t''}$ and $G_{X''t''}$ are of order $\rho'$, and $G_{X''X''}$ is of
order 1 as $r\to r_-$.
On the other hand, if $|\beta|>\gamma^2/2$, then it is possible to find a
$\phi$ ($\coth\phi=\beta (r_+-r_-)/2(r_--r_0)$) for which $G_{X''X''}$ and
$G_{X''t''}$ are of order $\rho'$ and $G_{t''t''}$ is of order unity as
$r\to r_-$.
Thus the global structure resembles that of a Reissner-Nordstrom black
hole in the first case, and that of the black string solution of
ref.\HORNE\TWIST\ in the second case.}
On the other hand, the point $r=r_0$ as well as $r=0$ represents genuine
singularities of
the solution.
($\Phi\to\pm\infty$ near these points.)
Since for real $\beta$ and $\gamma$, $r_0\le r_\pm$, we see that the
solution represents a genuine singularity surrounded by two horizons.
Solvable conformal field theories corresponding to black string solutions
with two horizons have been found previously by Horne and Horowitz\HORNE.

Note that if we take $\beta=0$, then the solution represents the direct
product of a four dimensional black hole carrying magnetic and electric
charge, and a six dimensional flat space described by the coordinates
$X^i$ ($1\le i\le 6$).
If we compactify the coordinates $X^i$ (say on a Calabi-Yau manifold, or a
six dimensional torus), the result would be a four dimensional black hole
carrying electric and magnetic charge.
(The full solution, on the other hand, may be regarded as a black string
in 5 dimensions by compactifying the coordinates $X^2,\ldots X^6$.)
The two charges, however, lie in different U(1) subgroups of the gauge
group.
These solutions  are different from the ones discussed in
ref.\WILCZEK\ in that in their solution the electric and the magnetic charge
lie in the same
U(1) subgroups of the gauge group.
On the other hand, these solutions can be identified to the black hole
solutions of
Gibbons and Maeda\MAEDA\ carrying electric and magnetic charge, if we
interprete the electric and magnetic charge in their solution to belong to
different U(1) subgroups of the gauge group.
(Note that this is the only way to interprete the solutions of Gibbons and
Maeda
in the context of string theory, since if the electric and the
magnetic fields belong to the same U(1) subgroup, we need to take
into account the effect of the gauge Chern Simons term coupling to the
antisymmetric tensor gauge field strength, which was not included in the
analysis of ref.\MAEDA.)
If we further set the magnetic charge $Q_M$ to zero, the solution reduces
to the charged black hole solution of ref.\GHS.
(Note that the metric $\hat{ds^2}$ considered in refs.\MAEDA\GHS\ is
related to the metric $ds^2$
given in eq.\etwelve\ through the relation $\hat{ds^2}=e^{-\Phi}
ds^2$\HOST.)

Since we have derived the transformation laws of various fields under
$\odp$ transformation even when $G_{m\alpha}$, $B_{m\alpha}$ and
$A_{\alpha R}$ are non zero, we could, in principle, perform an $\odp$
rotation  that includes the 1 direction of the gauge field.
Note, however, that in this case, the initial gauge potential needs to be
defined in separate coordinate patches; and are related by a gauge
transformation on the overlap.
This, in general, implies that the transformed fields also need to be
defined in separate coordinate patches, and are related by gauge and
general coordinate transformation on the overlap.
To see this let us consider the transformation of the fields in the
asymptotic region  $r\to\infty$, so that
$G_{\mu\nu}-\eta_{\mu\nu}$, $B_{\mu\nu}$ and $A_{\mu R}$ are small.
If we choose $S=1$, and $R$ to be a $O(1,1)$ transformation that mixes
the $t$ coordinate with the 1 direction in the internal space, the
transformed fields take the form:
$$\eqalign{
G'_{\alpha t} &= {1\over 2}\sinh\theta A^1_\alpha\cr
B'_{\alpha t} &= {1\over 2}\sinh\theta A^1_{\alpha}\cr
A^{1\prime}_{\alpha} &= A^1_\alpha \cosh\theta\cr
}
\eqn\egaugeone
$$
where $\alpha$ denotes any of the three directions $x$, $y$ or $z$
on which the original solution depends.
Let $A^1_\alpha$ and $\bar A^1_\alpha$ be the components of the original
gauge field in the two different coordinate patches, then $A^1_\alpha
-\bar A^1_\alpha= \p_\alpha\Lambda$, where $\Lambda$ is a function which
is not single valued under a $2\pi$ rotation about the $z$ axis (although
$e^{ie\Lambda}$ is).
{}From eq.\egaugeone\ we see that $G'_{\alpha t} -\bar G'_{\alpha t}$ is
now given by $(1/2)\sinh\theta\p_\alpha\Lambda$, where $G'$ and $\bar
G'$ denote the transformed  metric in the two coordinate patches.
This shows that $G'$ and $\bar G'$  are related by a coordinate
transformation of the form $t \to t + \Lambda\sinh\theta$.
However, since $\Lambda$ is not a single valued function of the
coordinates, this coordinate transformation is not globally well defined.

We could also have started with the metric which represents
black holes carrying quantised antisymmetric tensor gauge field
charge, instead of magnetic charge.
This solution is given by\HOST:
$$\eqalign{
ds^2 = & -{1-r_+^2/r^2\over 1-r_-^2/r^2} dt^2 +{dr^2\over (1-r_+^2/r^2)(
1-r_-^2/r^2)} + r^2d\Omega_3^2 +\sum_{i=1}^5 dX^i dX^i\cr
\Phi =& -\ln (1-r_-^2/r^2) +\Phi_0\cr
\tilde H_{\alpha\beta\gamma} =& Q(\epsilon_3)_{\alpha\beta\gamma}\cr
}
\eqn\eoriginalantisymm
$$
where $d\Omega_3$ is the line element on a three sphere, $\epsilon_3$ is
the volume form on the same three sphere, and $Q= r_+r_-$.
In this solution, the expressions for $G_{tt}$ and
$\Phi$ are similar to those given in eqs.\eseven\ and \eeight, except that
the ratios $r/r_\pm$ are replaced by $(r/r_\pm)^2$.
As a result, the final transformed solution will have the same form as
given in eqs.\etwelve\ and \ethirteen\ with $r$, $r_0$ and $r_\pm$
replaced by $r^2$, $(r_0)^2$ and $(r_\pm)^2$ everywhere in the expression
for $\hat G$, $\hat B$ and $\hat A$.
Thus the final solution will take the form:
$$\eqalign{
ds^2 = & -{1\over 4(r^2-r_0^2)^2}(4(r^2-r_+^2)(r^2-r_-^2)-(r_+^2 -
r_-^2)\beta^2) dt^2 + \beta{r_+^2 - r_-^2\over r^2-r_0^2} dX^1 dt
\cr
& +\sum_{i=1}^5
dX^idX^i + {dr^2\over (1-r_+^2/r^2)(1-r_-^2/r^2)} + r^2 d\Omega_3^2\cr
\hat B_{t1} =& \beta {r_+^2 - r_-^2\over 2(r^2-r_0^2)}\cr
A_t = & \gamma {r_+^2-r_-^2\over r^2 - r_0^2}\cr
A_1 = & 0\cr
\Phi = & -\ln(1-r_0^2/r^2)\cr
\tilde H_{\alpha\beta\gamma} = & Q(\epsilon_3)_{\alpha\beta\gamma}\cr
}
\eqn\eantisymonopole
$$
where,
$$
r_0^2 = {1\over 2} ((r_+^2 + r_-^2) - (r_+^2 - r_-^2)\sqrt{1
+\beta^2+\gamma^2})
\eqn\eantisytwo
$$
Note that in this case the antisymmetric tensor gauge field has a
`magnetic' type component denoted by $\tilde H_{\alpha\beta\gamma}$ and
also an electric type component denoted by $\hat H_{rt1}\equiv\p_r\hat
B_{t1}$.
Again, by taking the directions $X^2,\ldots X^5$ to be compact, this
solution may be regarded as a black string solution in six dimensions.

In some cases, one can get solvable conformal field theories describing
black hole solutions\con\WITTEN\MANDAL\EFR\BARDACKI\ROCEK\BANE\noc\DVV.
One expects that by twisting these solutions one will get solutions that
again correspond to solvable conformal field theories.
In fact the black p-brane solution obtained by twisting the
solution\SCALE\
given in ref\WITTEN\ are also described by solvable conformal field
theories\LISTEIF\HORNE.

\chapter{CONCLUSION}

In this paper we have shown that given a classical solution of the
heterotic string theory which is independent of $d$ of the space-time
coordinates, and for which the background gauge field lies in a subgroup
that commutes with $p$ of the $U(1)$ generators of the gauge group, we can
generate other classical solutions by applying an $\odp$ transformation on
the original solution.
By using these transformations on the known black 6-brane solution of the
heterotic string theory carrying magnetic charge, we have generated new
solutions carrying magnetic, electric and antisymmtric tensor gauge field
charge.
These solutions are labelled by four continuous and one discrete
parameters, characterizing
the mass, the electric charge, the antisymmetric tensor gauge field
charge, the
asymptotic value of the dilaton field, and the magnetic charge of the
6-brane respectively.
By compactifying 5 of the directions this solution may be regarded as a
black string solution in five dimensions.
Using this method we have also constructed black string solutions in six
dimensions carrying electric charge, and both, electric and magnetic type
antisymmetric tensor gauge field charge.

\refout
\end